\documentclass[12pt,preprint]{aastex}

\shorttitle{Near Infrared Background}
\shortauthors{Thompson et al.}

\begin{document}

\title{Evidence for a $Z<8$ Origin of the Source Subtracted Near
Infrared Background}

\author{Rodger I. Thompson, Daniel Eisenstein, Xiaohui Fan and Marcia Rieke}
\affil{Steward Observatory, University of Arizona,
    Tucson, AZ 85721}
\email{rthompson@as.arizona.edu, deisenstein@as.arizona.edu, fan@as.arizona.edu, mrieke@as.arizona.edu}

\author{Robert C. Kennicutt\altaffilmark{1}}
\affil{Institute of Astronomy, University of Cambridge, 
Cambridge CB3 OHA UK}
\email{robk@ast.cam.ac.uk}

\altaffiltext{1}{Steward Observatory, University of Arizona,
    Tucson, AZ 85721}

\begin{abstract}

This letter extends our previous fluctuation analysis of the near infrared 
background at 1.6$\micron$ to the 1.1$\micron$ (F110W) image of the Hubble 
Ultra Deep field.  When all detectable sources are removed the ratio of 
fluctuation power in the two images is consistent with the ratio expected 
for faint, $z<8$, sources, and is inconsistent with the expected ratio 
for galaxies with $z>8$. We also use numerically redshifted model galaxy 
spectral energy distributions for 50 and 10 million year old galaxies to 
predict the expected fluctuation power at 3.6$\micron$ and 4.5$\micron$ 
to compare with recent \emph{Spitzer} observations.  The predicted 
fluctuation power for galaxies at z = 0--12 matches the observed 
\emph{Spitzer} fluctuation power while the predicted power for $z>13$ 
galaxies is much higher than the observed values.  As was found in the 
1.6$\micron$ (F160W) analysis the fluctuation power in the source subtracted 
F110W image is two orders of magnitude below the power in the image with all 
sources present.  This leads to the conclusion that the 0.8--1.8$\micron$ near 
infrared background is due to resolved galaxies in the redshift range $z<8$,
with the majority of power in the redshift range of 0.5--1.5.  

\end{abstract}


\section{Introduction}

In a previous paper \citep{thm07} we addressed the nature of the Near InfraRed
Background (NIRB) at 1.6$\micron$ using NICMOS observations in the Hubble Ultra
Deep Field (HUDF).  The area covered by the NICMOS observations comprises the 
NICMOS Ultra Deep Field (NUDF) which is smaller than the HUDF.  The NIRB at 
1.1$\micron$ and 1.6$\micron$ has a total power of 7 nWm$^{-2}$sr$^{-1}$ emitted 
by resolved galaxies, predominantly in the redshift range z = 0.5--1.5.  This is 
in contrast to the previous results from \citet{mat05} that claimed a peak flux
of 70 nWm$^{-2}$sr$^{-1}$ at 1.4$\micron$.  The discrepancy arose not in the 
value of the total measured flux but in the attribution of the flux to the
components of zodiacal emission, resolved stars and galaxies and an excess.
In \citet{mat05} models were used to determine the first two components and
the difference between the modeled components and the total observed flux
was attributed to an excess.  The \citet{mat05} measurements did not extend
shortward of 1.4$\micron$ and the sharp drop in the flux levels between the 
1.4$\micron$ value and shorter wavelengths was initially attributed by some
authors (eg. \citet{sal03}) as due to the Lyman break in very high redshift
galaxies.  Our analysis measured the flux from the resolved stars and galaxies
and the zodiacal background and found no excess.  The difference was in the
amount of flux attributed to the zodiacal background.  Our measured value was
higher than the modeled value used by \citet{mat05} and the difference was
equal to the flux attributed to an excess.  We therefore concluded that unless
there was a extragalactic flux component that was extremely flat, mimicking 
zodiacal flux, there is no excess and the near infrared background is 
resolved.  

\citet{thm07} also addressed the spatial fluctuations in the NUDF field relative
to the fluctuations found in deep calibration 2MASS images at the same wavelength
by \citet{kas02}. \citet{kas02} subtracted all of the resolved sources from
the 2MASS images and found fluctuations in excess of that expected by shot
noise at long wavelengths. The excess was attributed to fluctuation power from 
galaxies at redshifts of 10 and above in several studies (eg. \citet{mag03}). 
A fluctuation analysis of the NUDF 
F160W image showed that when sources were removed from the NICMOS image down to 
the level that were removed from the 2MASS image, the fluctuations were consistent 
with those found by \citet{kas02}.  When sources were removed down to the
detection limit of the combined ACS and NICMOS observations in the NUDF the
fluctuation power dropped by almost two orders of magnitude.  Since all of the
detected sources had redshifts between 0 and 8, this meant that the fluctuations
found by \citet{kas02} were from galaxies in the redshift range between 0 and 8.
Further analysis in \citet{thm07} of images that only retained sources in a given
redshift bin determined that the majority of fluctuation power came from galaxies
in redshift range between 0.5 and 1.5. The fluctuation spectra from
galaxies with redshifts greater than 4 were not detectable above the noise. 

More recent fluctuation analyses of \emph{Spitzer} IRAC images at 3.6$\micron$ 
and 4.5$\micron$ (\citet{kas05b} and \citet{kas07b}) interpret the residual 
fluctuations after source subtraction as being due to high redshift, possibly 
Population III, galaxies \citep{kas07a}. This assertion appears to be based
on the assumption that all galaxies below the IRAC detection limit must be faint
due to their high redshift. This interpretation has been challenged by 
\citet{coo07} who did further source subtraction in the field using deep ACS
images to identify faint sources missed by the IRAC images.  They found
the removal of these sources greatly reduced the signal seen by \citet{kas07b}
and the signal must be due to sources with redshifts less than 8 that are
visible in the ACS images. \citet{kas07a} replied that the excessive amount of
area removed from the IRAC images in the source subtraction carried out by
\citet{coo07} invalidates their result. Modeling by \citet{sal06} indicates 
that the observed IRAC fluctuations may be consistent with those created by
Pop II stars at redshift greater than or equal to 5. The present study is 
motivated in part by a desire to investigate whether the observations in 
the NUDF at both 1.1$\micron$ and 1.6$\micron$ are consistent with the high 
redshift interpretation of the the IRAC observations by examining the color 
of the fluctuations and their expected extrapolation to the IRAC bands.

\section{Data and Analysis}

The location, size and pixel scale of the F110W image is identical to the 
F160W NUDF image analyzed in \citet{thm07}. Details of the image preparation
are given in \citet{thm05}.  The basic image production after the processing
of the individual images is production of a background image which is the
median of all of the individual images, subtraction of the background from
the individual images and then combining the images with the drizzle procedure
\citep{fru02}. The fluctuation analysis of the F110W image is identical to the 
analysis techniques used on the F160W image in \citet{thm07} and is not repeated 
in detail here. Source subtraction is accomplished using the SExtractor 
\citep{ber96} (SE) pixel map where each pixel that is part of a source is given
a value equal to the source ID number and all other pixels have a value of 
zero.  The photometric redshifts derived in \citet{thm06} were used to identify
the redshift of each source.  Due to the much higher resolution of the NICMOS
images relative to the IRAC images, even the all source subtracted image retained
93$\%$ of its pixels.  This means that the objections of \citet{kas07a} to 
the \citet{coo07} analysis do not apply here.  All source pixels were set to
zero in the subtracted image.  There was no attempt to replace them with random
noise.  The small area of the subtracted sources means that this had no affect
on the final fluctuation spectrum. Figure~\ref{fig-fluct} shows the fluctuation 
power at 1.1$\micron$ and 1.6$\micron$. To determine the redshift distribution of 
the 1.1$\micron$ fluctuations the fluctuation analysis was also performed on the 
1.1$\micron$ image with all sources except those in a given redshift bin removed. 
The 1.1$\micron$ fluctuation power versus 
redshift distribution is essentially identical to the 1.6$\micron$ distribution 
shown in Figure 6 of \citet{thm07}. There is a peak near z=1 and no discernible 
power above the background for $z>4$. 
 
\section{Fluctuation Colors \label{s-fc}}

To investigate the nature of the sources of the fluctuation power observed in
the all source subtracted NICMOS and IRAC images we calculate
the expected fluctuation power color as a function of redshift using galaxy
Spectral Energy Distributions (SEDs). At any given angular scale the fluctuation 
power for a given wavelength band is directly proportional to the power of the 
galaxies ($\nu I_{\nu}$) in that band.  The power ratios  of the bands at any 
redshift are easily calculated from the numerically redshifted model SEDs. 
The analysis of \citet{thm06} used 7 primary template SEDs labeled 1--7 going
from early to late galaxies.  The first 5 are observed SEDs and SEDs 6 and
7 are calculated from the models of \citet{bru03}.  The properties of these
models are given in Table~\ref{tab-temp}.  Template 6 is for a 50 million 
year old galaxy and Template 7 is for a 10 million year old galaxy with the 
lowest metallicity available in the \citet{bru03} models.  Template 7 is 
similar to a Pop. III SED but redder than a true Pop. III SED (see 
\citet{sch02} for the expected differences).  Although a Pop. III SED may be 
bluer than template 7, the sharp increase in the 3.6$\micron$ to 1.6$\micron$ 
fluctuation power at high redshift is due to the Lyman break passing through 
the 1.6$\micron$ and is therefore relatively independent of the SED to the 
red of the break.    

\citet{thm06} interpolated 
between the SEDs in increments of 0.1 to produce a final set of 61 different
model SEDs. Figure~\ref{fig-temp} shows the distribution of the template
numbers for all sources in the NUDF with redshifts greater than 2.0.  The
predominance of sources concentrate on templates 6 and 7, therefore, we will take
them as the most probable SEDs for sources fainter than our source 
detection limit.  We use these SEDs and the effective response functions 
in the NICMOS and IRAC filters to calculate the expected ratios of the power at 
redshifts between 0 and 15.  Absorption in the Inter-Galactic Medium is modeled
according to the treatment given in \citet{mad96}.  

The horizontal lines in Figure~\ref{fig-col} show the observed ratios of 
the fluctuation power in the all source subtracted images using; this analysis, 
\citet{kas07b} and \citet{thm07}. Since the excess fluctuation power attributed 
to high redshift galaxies by \citet{kas07b} is at angular scales greater than 
$30\arcsec$ the horizontal lines show the ratios observed at $100\arcsec$, the 
largest NICMOS angular scale.

The observed 1.1$\micron$ to 1.6$\micron$ fluctuation power ratio is consistent 
with the lower redshift template 6 predictions, but is incompatible with the 
calculated ratios at $z>8$ for either of the templates.  This is strong 
evidence that the source subtracted fluctuations in the NICMOS images are 
due to galaxies at $z<8$. The observed 3.6$\micron$ to 1.6$\micron$ ratio 
lies between the ratios predicted by templates 6 and 7. Given the good match 
to the NICMOS flux ratios for template 6 we would expect higher 3.6$\micron$ 
fluctuations than observed.  The difference may be due to cosmic variance since 
the fields are different.  It may also be that the much broader \emph{Spitzer} 
PSFs subtract more background flux than the HST PSFs. For $z>12$ the 
predicted ratios become incompatible with observations. The 4.5$\micron$ to 
3.6$\micron$ ratio is essentially independent of redshift at the redshifts 
considered and therefore does not contribute any information on the redshift 
epoch of the fluctuations.  The apparent agreement between the predicted and 
observed ratios for those bands, however, is further evidence that the observed 
fluctuations are due to faint galaxies at redshifts less than 8.

\subsection{Compatibility of the Data Sets}

An important question when comparing the NICMOS and IRAC source subtracted remnant 
fluctuations is whether sources are subtracted to the same level in both data sets.
Although \citet{kas07b} does not explicitly state the achieved source subtraction 
level, \citet{kas07a} states that any remaining sources in the IRAC image must be
fainter than 10--20 nJy.  For the SEDs used in this study the IRAC limit 
corresponds to a NICMOS limit of 4.5--9 nJy.  From Figure 3b from \citet{thm07} 
the completeness limit is approximately 20 nJy at 1.6$\micron$, which is 2 to 4
times brighter than the IRAC subtraction limit.  The sources for subtraction,
however, were identified using detections in a combination of the ACS and 
NICMOS HUDF
images, with the ACS images going significantly deeper due to their much longer
integration time.  Figure 7 in \citet{thm06} shows that the detection limit in
the ACS F775W band is approximately 30 in AB magnitudes which is 3.7 nJy.  We can
use the SEDs discussed in \S\ref{s-fc} to see what IRAC fluxes these correspond
to.  The results are shown Figure~\ref{fig-cds}.  Template SED 6 gives a limit of
20 nJy and the bluer template 7 SED gives a limit of 10 nJy which are compatible
with the IRAC subtraction limits given by \citet{kas07a}.  The F775W band only 
removes sources at that limit for redshifts less than 3 but the ACS F850LP long
pass filter removes sources at that limit for redshifts up to 6.  The two data
sets are compatible for all redshifts less than 6.  At redshifts between 6 and 8
the NICMOS subtraction limit is not as stringent as the IRAC subtraction limit.
To account for this difference the horizontal solid line in Figures~\ref{fig-col}a
and b would have to be raised by a factor of 2 to 4 but that would still not be
compatible with redshifts greater than 12.  The redshift distribution of galaxies
with NICMOS fluxes less than 20 nJy would have to be significantly different
than galaxies above this limit for the z = 6--8 sources to contribute any power
to the observed NICMOS fluctuations.  

\subsection{Signal or Noise?}

\begin{figure}
\plottwo{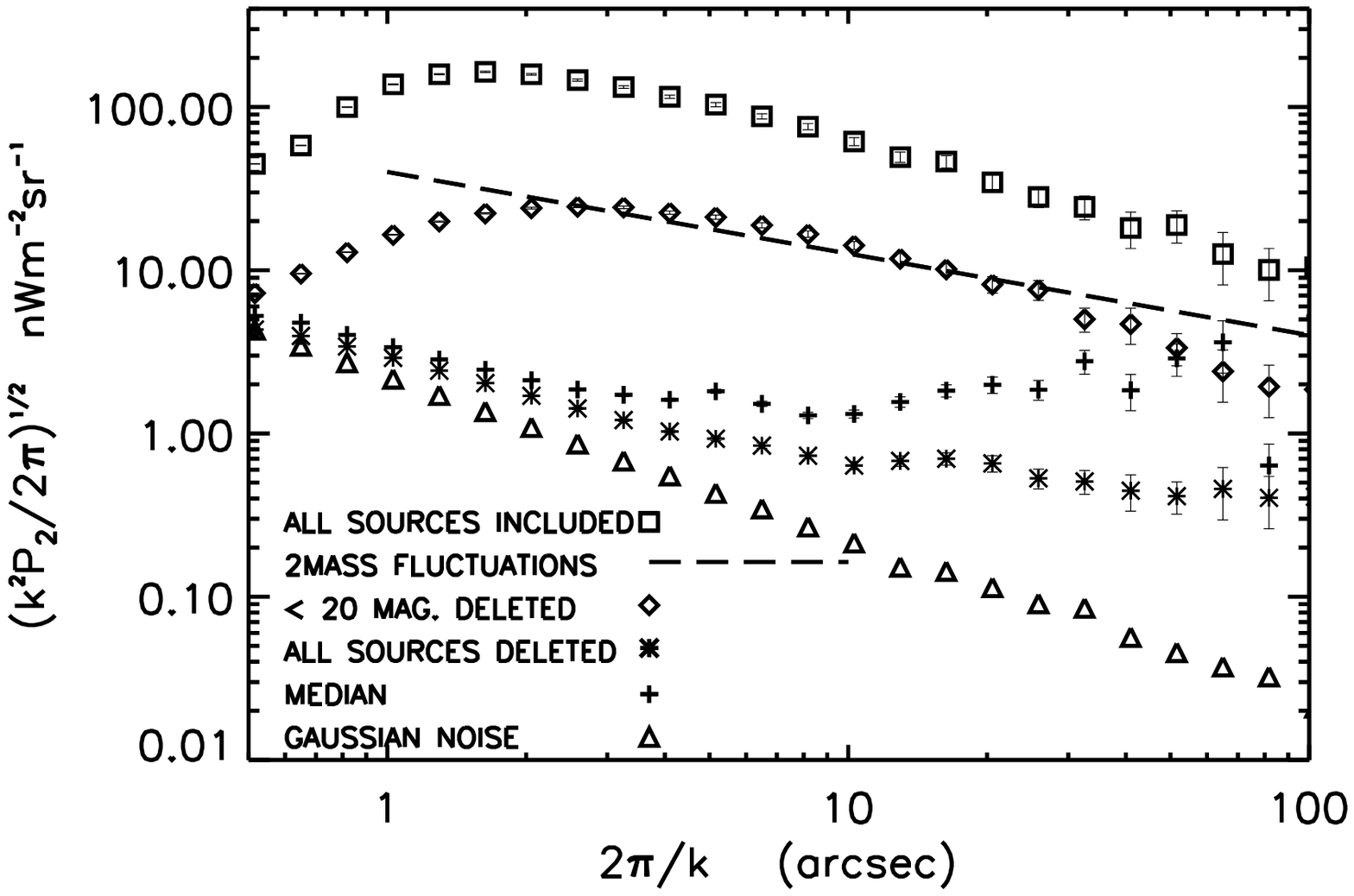}{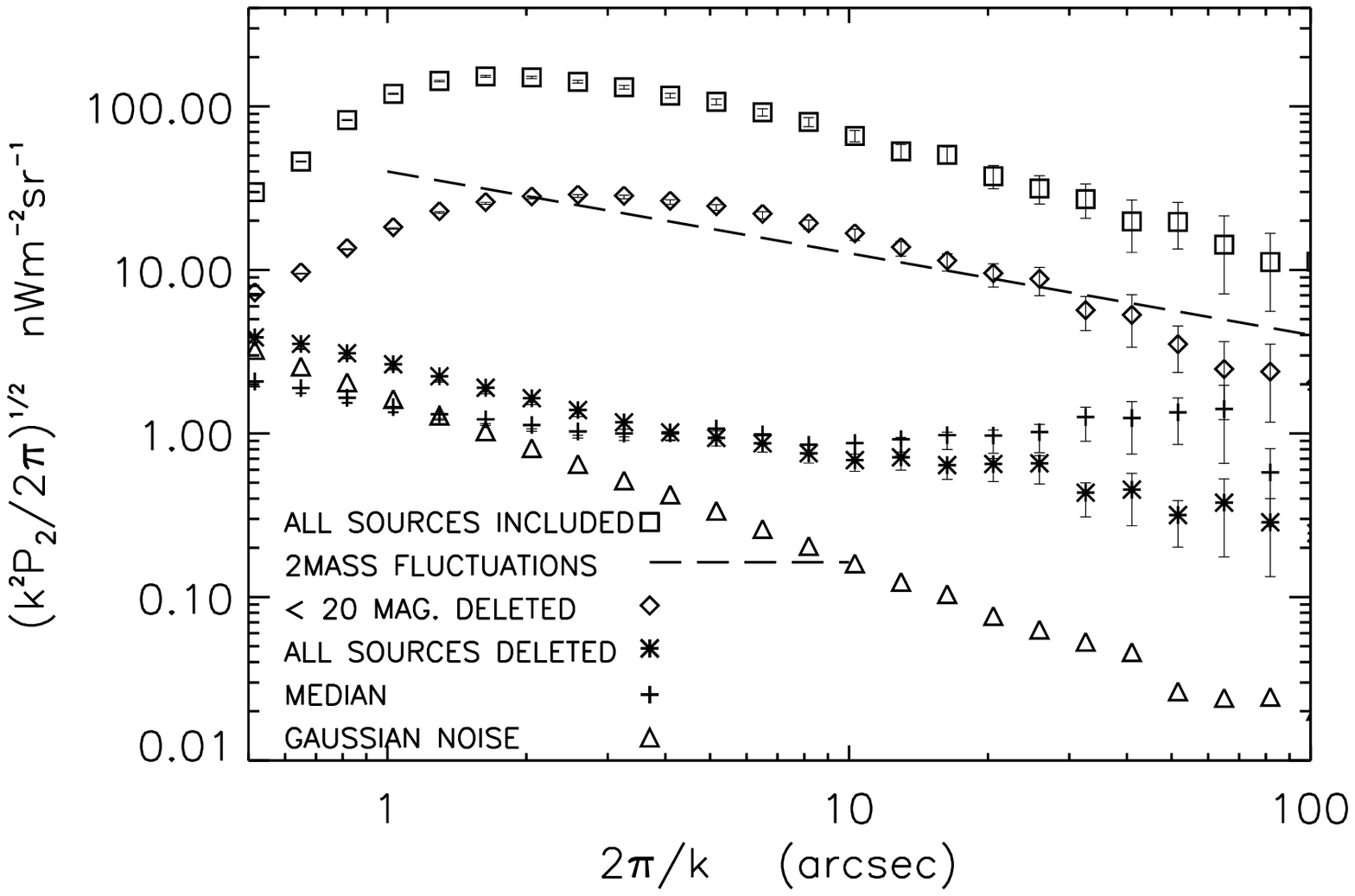}
\caption{The fluctuation spectra from a) the F110W NUDF image (left) and b) the F160W
image (right).  The open squares are for no sources subtracted, the open diamonds
are for all sources with a F160W AB magnitude of 20 or brighter subtracted. 
The asterisks are for all sources subtracted and the crosses are for the
median background field described in the text.  The open triangles are 
for a Gaussian distributed noise image with a standard deviation equal to 
that of the images. The dashed line indicates the level of 1.6$\micron$ 
fluctuations found in deep calibration 2MASS fields by \citet{kas02}.
\label{fig-fluct}}
\end{figure}

\begin{figure}
\plotone{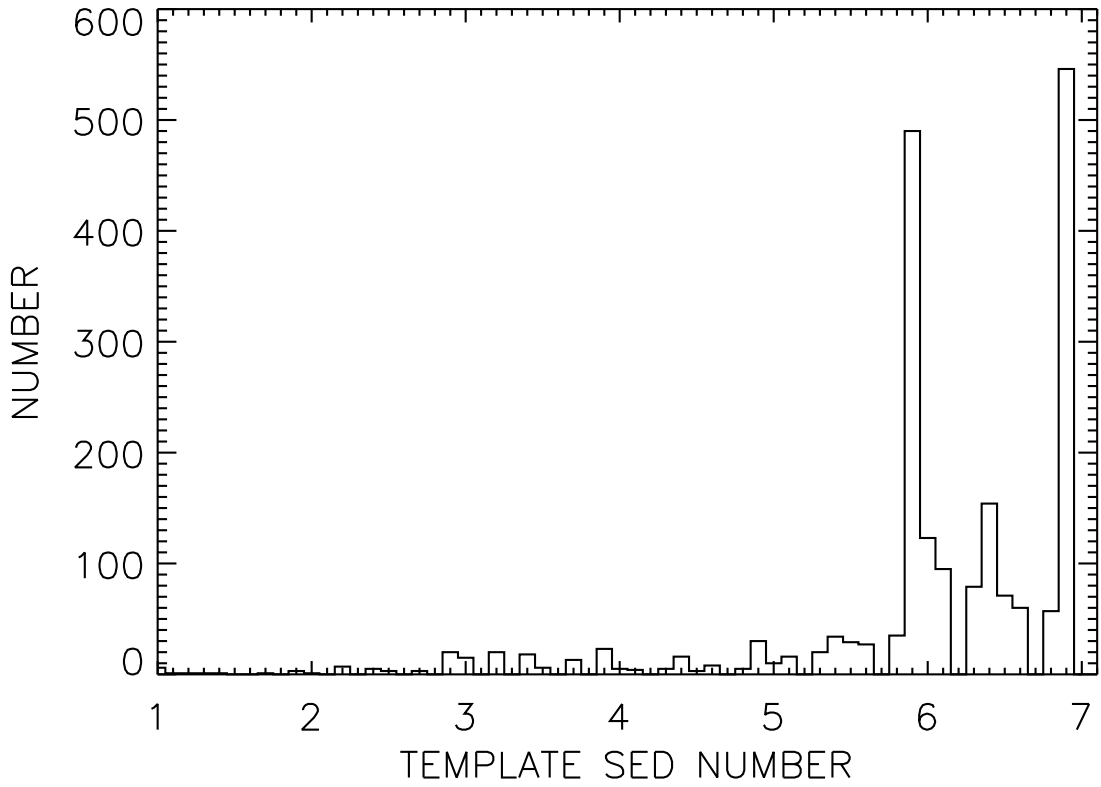}
\caption{The histogram of SED template numbers for all sources with redshifts
greater than 2.0.  The interpolated templates go in increments of 0.1 between
the primary template SEDs.
\label{fig-temp}}
\end{figure}

\begin{figure}
\plottwo{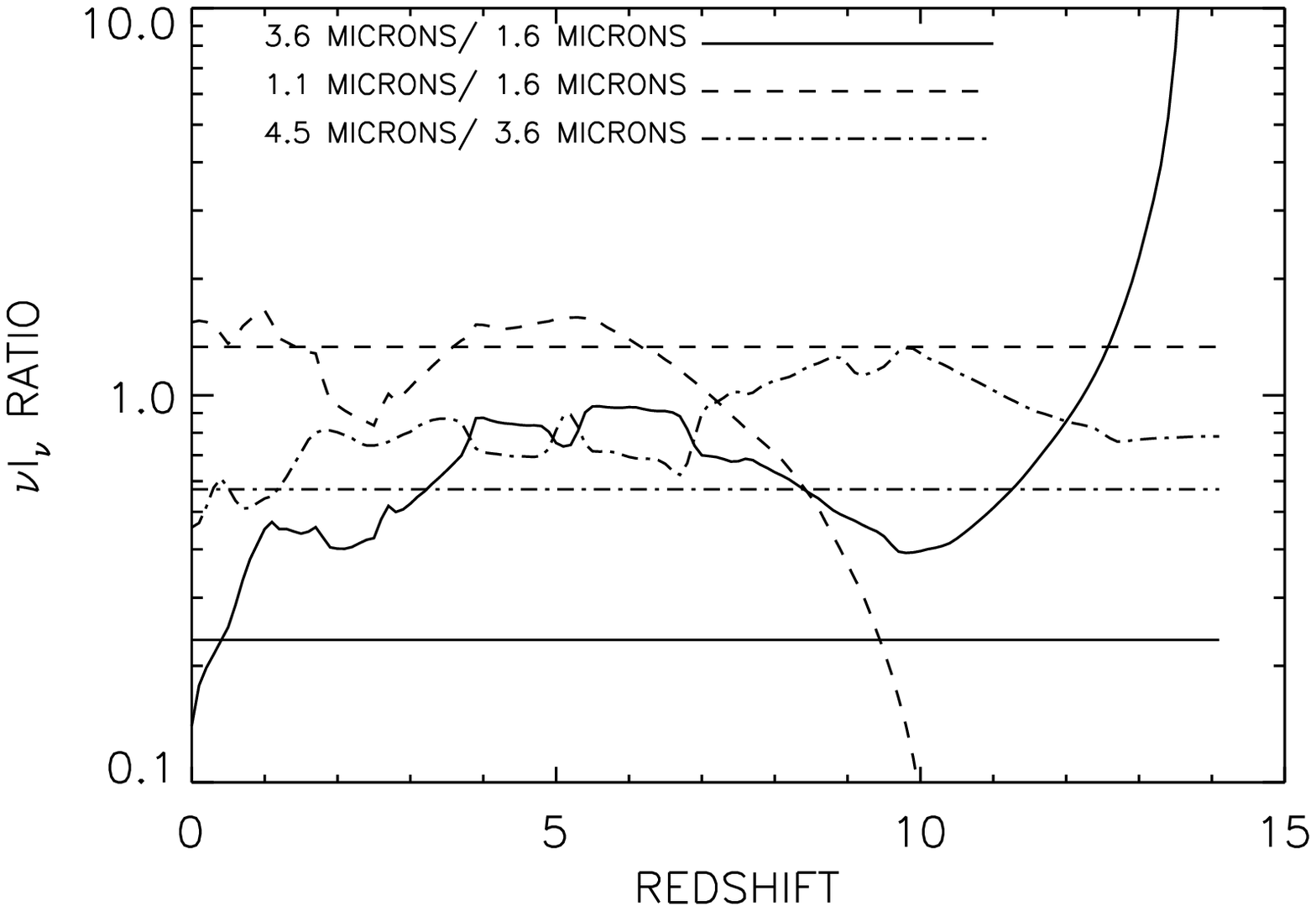}{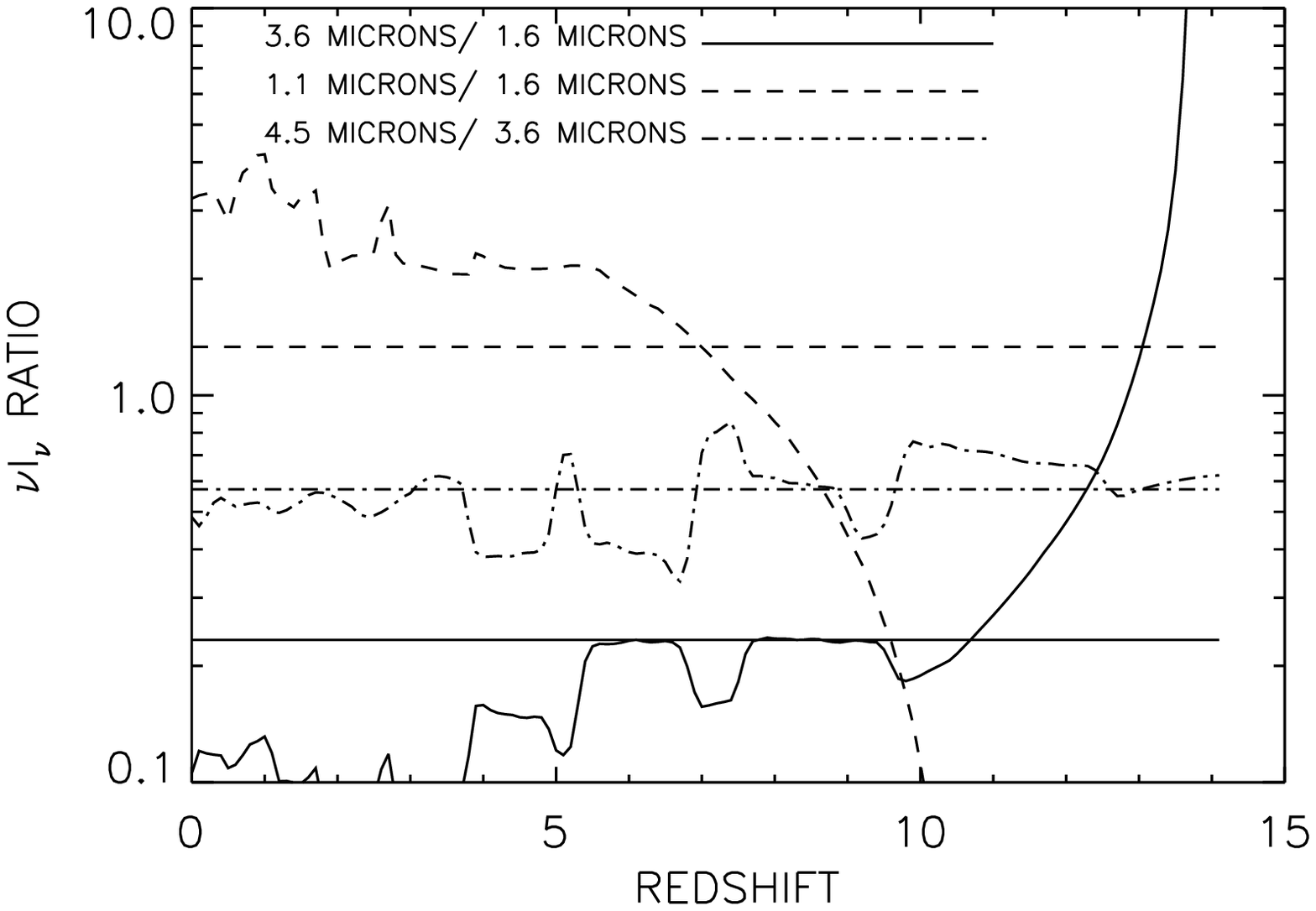}
\caption{The predicted ratios of the powers, $\nu I_{\nu}$, in the F110W band 
to the F160W band, dashed line; the the 3.6$\micron$ \emph{Spitzer} band  to 
F160W band, solid line; and the \emph{Spitzer} 4.5$\micron$ to 3.6$\micron$  
band, dash dot line; versus redshift. The straight horizontal lines indicate 
the observed values of the ratios at an angular scale of $100\arcsec$ with the 
same line style code. a) (left) is for template6 and b) (right) is for 
template 7, the hottest SED, which is consistent with a possible 
Population III galaxy.
\label{fig-col}}
\end{figure}

\begin{figure}
\plotone{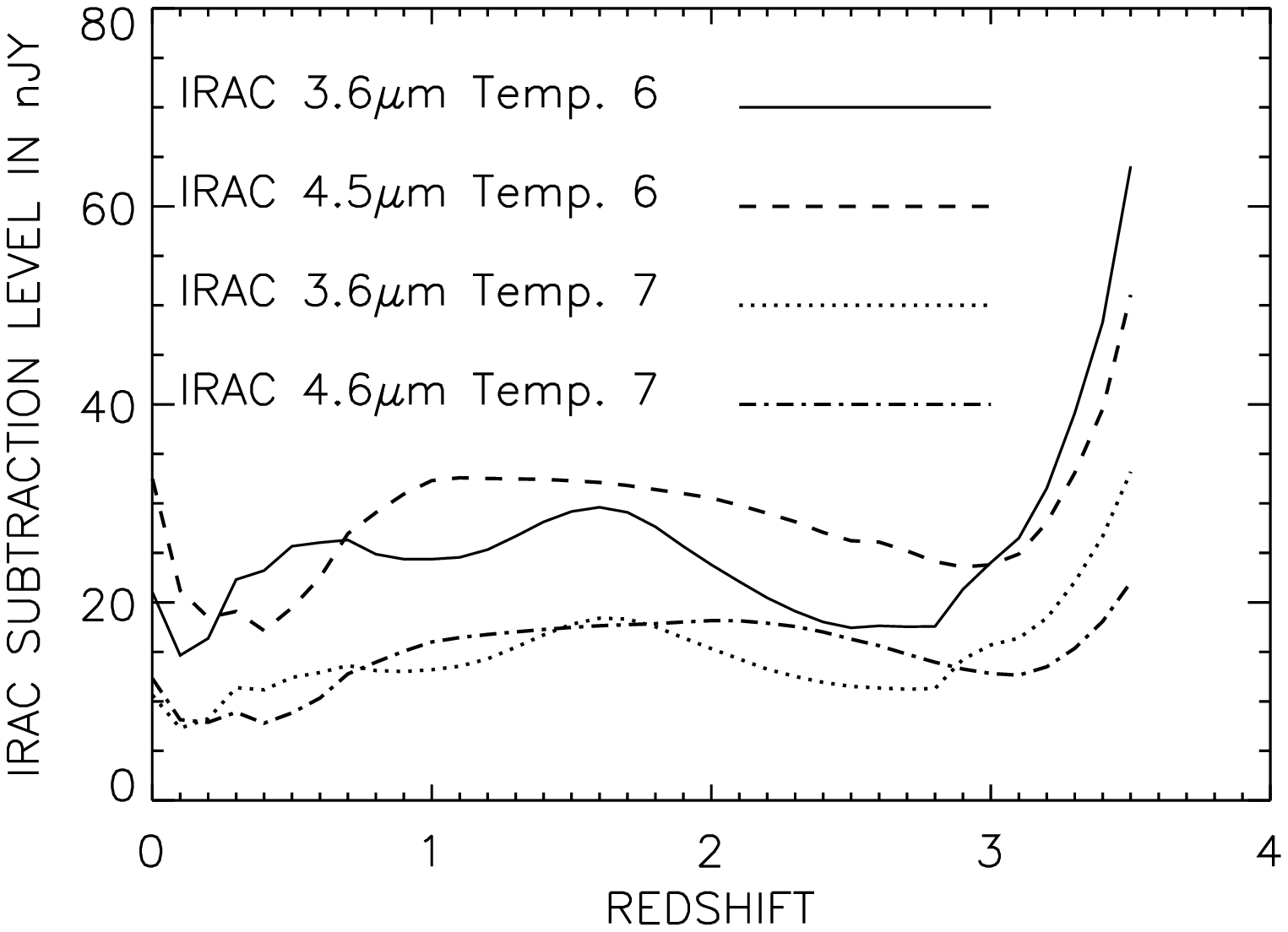}
\caption{The figure shows the IRAC fluxes in the 3.6$\micron$ and 4.5$\micron$
bands that correspond to a subtraction limit of 3.7 nJy in the ACS F775W band
as a function of wavelength for the two templates (6 and 7) used in this study.
The quoted subtraction level from \citet{kas07b} is 10 to 20 nJy, consistent 
with subtraction levels in the figure.  The sharp rise at approximately z=3 
is when the Lyman break starts to enter the F775W band.
\label{fig-cds}}
\end{figure}

All of the discussion above is predicated on the assumption that the fluctuations
observed after source subtraction are due to faint galaxies below the detection limit
and not noise sources from the instrument or telescope. To check for the signature of
Gaussian distributed noise in the NICMOS images we populated an image of the same
size as the UDF F110W and F160W images with random Gaussian distributed noise with
a variance equal to that found in the real images.  An identical fluctuation
analysis gives the results shown by the triangles
in Figure~\ref{fig-fluct}.  The random noise is close to the observed fluctuation power
at small angular scales, but is far below the observed power at angular scales of 
$10\arcsec$
and larger, the region considered in this analysis.  We therefore conclude that
the observed fluctuation power is not due to random noise components such as read noise,
Poisson photon statistics, or dark current in the NICMOS images.

We originally did not include the F110W image in the fluctuation analysis due to the 
possibility of residual flat fielding errors.  The median of the 144 individual images 
measures any flat field errors very accurately.  The F110W mosaic image created by 
replacing all of the individual actual images with the median image does appear to 
contain a component of noise due to small flat field errors.  It is evident as the 
higher fluctuation power in Figure~\ref{fig-fluct}a at spatial scales between
$10\arcsec$ and $80\arcsec$.  A similar signature is not present in the 1.1$\micron$ 
source subtracted fluctuation power.  This indicates that the subtraction of the median 
image from each actual image before the mosaic accurately removes any flat field
errors in the regions that have no detectable sources.  These are the regions that
contribute to the all source subtracted fluctuation power.  We therefore exclude flat 
fielding errors as a source of fluctuation power.

\begin{figure}
\plottwo{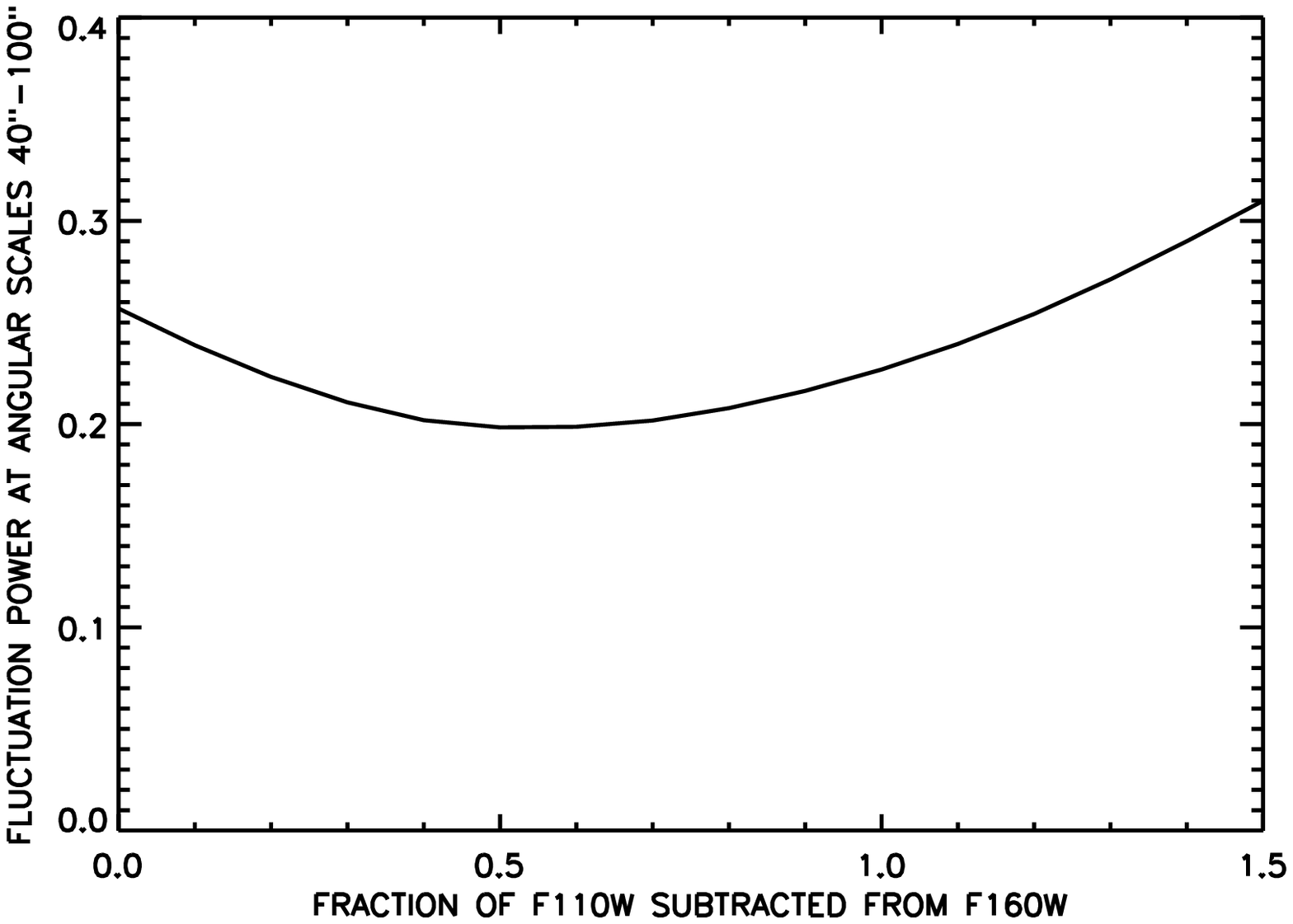}{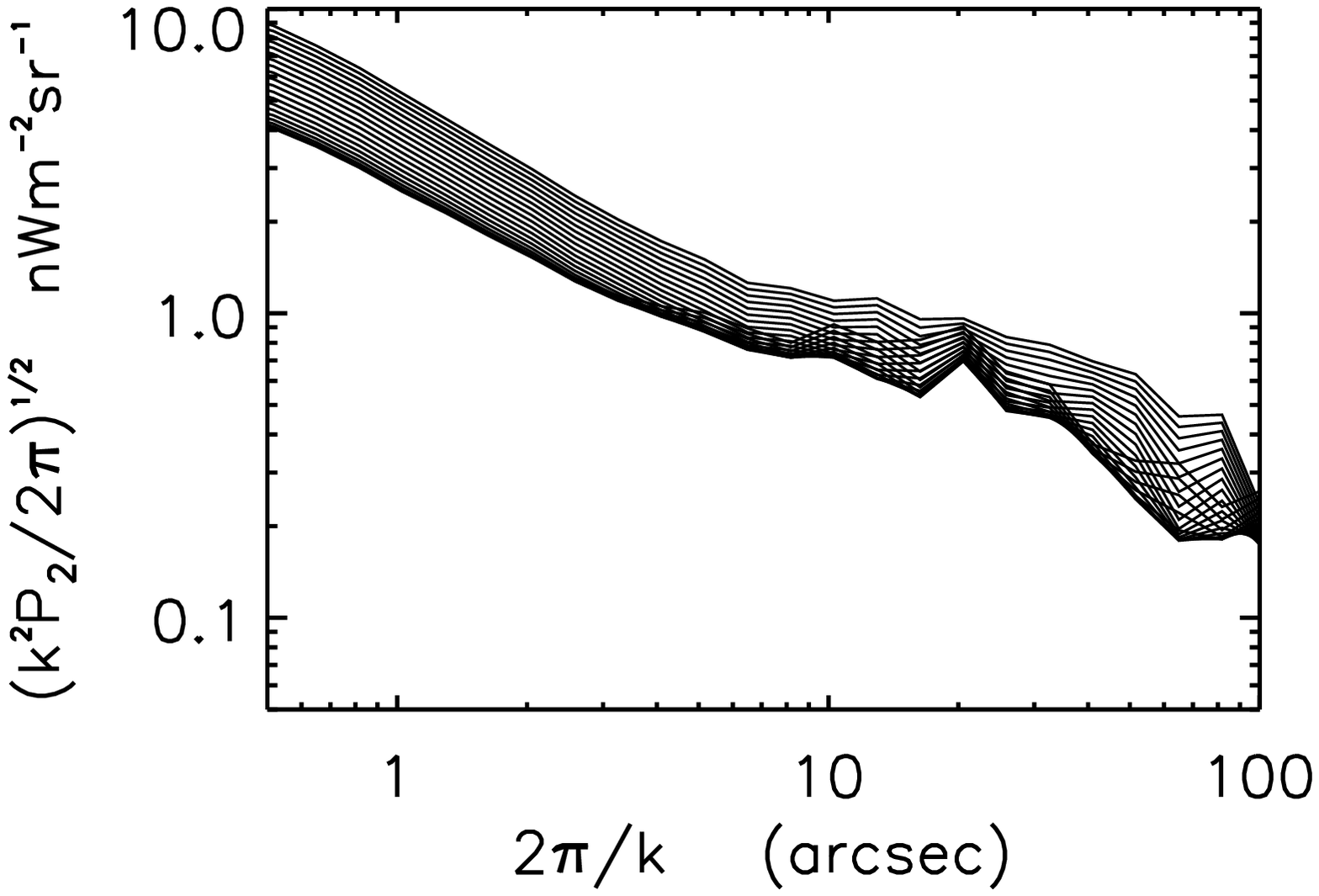}
\caption{a) The left panels shows the $\ln(k)$ weighted sum of fluctuation 
power for angular scales greater
than $30\arcsec$ versus the scaling factor of the F110W image subtracted from the
F160W image.  The minimum is at a scaling factor of 0.6. b) The right panel shows the nested
fluctuation power spectra for the different scaling factors.  The lower boundary is the 
limit for the contribution of high redshift sources plus other sources of spatial noise.
\label{fig-sub}}
\end{figure}

If the source subtracted fluctuations are due to faint, low redshift, galaxies then 
the 1.1$\micron$ and 1.6$\micron$ source subtracted images should be spatially correlated.  
A simple check for correlation is to subtract the F110W image from the F160W to test 
whether the fluctuations are reduced as would be expected if they are correlated.
Since the exact scaling is not known we subtracted the F110W image with scalings 
between 0 and 2 in 0.1 increments.  The log weighted sum of the fluctuation powers 
for angular scales greater than $30\arcsec$ versus the scale factor is shown in 
Figure~\ref{fig-sub}a.  There is a broad minimum for a scale factor of 0.6, near 
the observed ratio of F160W to F110W fluctuation of 0.75 at $100\arcsec$. 
Figure~\ref{fig-sub}b shows the nested fluctuation spectra.  This 
shows that the subtraction of the scaled F110W image reduces and alters the fluctuation
spectrum at larger angular scales. It also indicates that the two images are partially
correlated after the subtraction of all known sources as would be expected if fainter 
undetected sources were present in the images.  The lower boundary in Figure~\ref{fig-sub}b
is the upper limit on fluctuation power contributed by high redshift, 1.1$\micron$ dropout, 
galaxies plus any remaining noise in the images.  This limit is still above the 
Gaussian noise spectrum shown in Figure~\ref{fig-fluct}a and b.

At the largest scale of $100\arcsec$ the residual fluctuation power in the F160W
minus F110W image is 0.2 nWm$^{-2}$sr$^{-1}$. If the residual is due to
sources with redshifts less than 10 then the expected power in the IRAC
3.6$\micron$ band is 0.02--0.2 nWm$^{-2}$sr$^{-1}$ depending on which SED
template is used.  This is consistent with the 0.1 nWm$^{-2}$sr$^{-1}$
power observed by \citet{kas07b} at the same spatial scale.  If the residual
power is due to galaxies at redshifts greater than 8 then the predicted
power is in the range of 0.04 to a very large number depending on the SED
and redshift.  The predicted power starts to exceed the observed flux at a
redshift of 11 for template 6 and at a redshift of 12 for template 7.  At
a redshift of 13.5 the predicted power exceeds the observed 3.6$\micron$ 
power by a factor of 10 for both SEDs.  The residual F160W minus F110W power
to 3.6$\micron$ power ratio is therefore consistent with emission from 
galaxies with redshifts less than 8 but cannot definitively rule out
that the observed 3.6$\micron$ fluctuation power comes from galaxies
in the redshift range between 8 and 13.  Redshifts beyond 13, however,
are not consistent with the ratio if the difference between the lower
limit in Figure~\ref{fig-sub}b and the Gaussian noise is due to real
sources.  It should be noted, however, that the predominant source of
fluctuation power in the observed sources comes from galaxies at 
redshifts between 0.5 and 1.5. Having the residual F160W minus F110W 
fluctuation power coming from galaxies with redshifts above 8 would 
require a remarkable suppression of the luminosity function of the 
0.5--1.5 redshift galaxies at low luminosities.

A final, but not as compelling, piece of evidence that the residual fluctuations are
due to galaxies is that the 1.1$\micron$ to 1.6$\micron$ fluctuation power color is 
similar to that expected from $z<8$ galaxies.  We cannot completely rule out 
an unknown source of spatial power, which would be the same in both images due to 
their similar observation pattern and data analysis.  However, in the absence of any 
known source of such spatial power we conclude that the 
NICMOS source subtracted fluctuations are due to faint, $z<8$, galaxies below the detection limit.
These galaxies become evident in the fluctuation analysis because it utilizes $93\%$ of 
the pixels while the detected source power is contributed by only $7\%$ of the image pixels.

Without far more knowledge of the IRAC instrument and data analysis procedures
than we currently possess we cannot make the same arguments for the faint galaxy
origin of the IRAC fluctuations.  We can say that the IRAC fluctuations are at 
the levels predicted for faint, $z<8$, galaxies from the NICMOS observations.  
Since all four bands have ratios consistent with a faint, $z<8$, galaxies 
origin we assume that the fluctuations in all four bands have extragalactic 
origins.  We note that even if the IRAC fluctuations were due to noise, it would 
still support our conclusions that the IRAC fluctuations are not due to very high 
redshift galaxies.

\section{Conclusions}

Our basic conclusion is that the source subtracted fluctuations observed in the two NICMOS bands
and the IRAC bands at 3.6$\micron$ and 4.5$\micron$ are consistent with faint,
$z<8$, galaxies that are below the individual detection limit in the
respective images.  We also conclude that the observed ratio of 1.1$\micron$ 
to 1.6$\micron$
fluctuations is incompatible with a significant fraction of power from $z>8$
galaxies and that the ratio of the IRAC 3.6$\micron$ to NICMOS 1.6$\micron$
fluctuation power is incompatible for galaxies at $z>13$.  In light of these 
observations we find no evidence that any of the observed fluctuations require 
flux from high redshift galaxies.  Of course such galaxies may exist, but  we 
conclude that they do not contribute measurable power to the current observations.

The argument in \citet{kas07a} for the IRAC fluctuations being due to high
redshift galaxies appears to be based on the assumption that any sources
fainter than their detection limit must be at high redshift because they are 
faint.  Examination of Figure 3b in \citet{thm07} shows that the z = 0--7
galaxy luminosity function shows no break up to a completeness 
limit of 5 nJy in the F160W band.  This implies that there are many more
low redshift galaxies fainter than the detection limit, i.e. galaxies that are not
faint because they are at $z>8$. 

A caveat relevant to this study is that if it is postulated that the IRAC 
fluctuations are due to galaxies with $z>13$ then the predicted 
1.6$\micron$ fluctuations would be below our detection level due to the 
very high ratio of IRAC power to NICMOS power at high redshifts.  However if 
these high-redshift galaxies are the only source of the IRAC fluctuations, 
then a mechanism must be found to suppress the low redshift IRAC fluctuations 
predicted from the NICMOS residual fluctuations.  At this time we know of no 
such mechanism.

\acknowledgments
This article is based on data from observations with the NASA/ESA Hubble Space Telescope
obtained at the Space Telescope Science Institute, which is operated by the 
Association of Universities for Research in Astronomy under NASA contract
NAS 5-26555.  We would like to thank the anonymous referee for helpful 
comments in the preparation of this article.

\clearpage

\clearpage

\clearpage

\clearpage

\clearpage

\clearpage

\clearpage

\begin{deluxetable}{lcccc}
\tabletypesize{\scriptsize}
\tablecaption{Properties of the template SEDs.  Both SEDs are taken from
models of \citet{bru03} based on the Padova 1994 models.  The IMFs are from 
\citet{cha03}.  \label{tab-temp}}
\tablewidth{0pt}
\tablehead{
\colhead{Template Number} & \colhead{Metallicity} & \colhead{IMF} & \colhead{Low Mass Limit} & \colhead{Upper Mass Limit}
}
\startdata
6 & 0.004 & Chabrier & 0.1 & 100\\
7 & 0.0001 & Chabrier & 0.1 & 100\\
\enddata

\end{deluxetable}

\end{document}